\documentclass[prl,twocolumn,showpacs,floatfix,amsfonts]{revtex4}
\usepackage{graphicx,graphics,color,epsfig} % Include figure files
\usepackage{bm}
\usepackage{amsmath}
\usepackage{amssymb}
\begin{document}

\title{The Fock-Darwin States of Dirac Electrons in Graphene-based
Artificial Atoms }
\author{Hong-Yi Chen$^*$, Vadim Apalkov$^\dagger$, and Tapash 
Chakraborty$^*$}
\affiliation{$^*$Department of Physics and Astronomy, University of 
Manitoba, Winnipeg, MB R2T 2N2, Canada \\
$^\dagger$Department of Physics and Astronomy, Georgia State 
University, Atlanta, Georgia 30303, USA}

\begin{abstract}
We have investigated the Fock-Darwin states of the massless 
chiral fermions confined in a graphitic parabolic quantum dot. 
In the light of the Klein tunneling, we have analyzed the
condition for confinement of the Dirac fermions in a 
cylindrically-symmetric potential. New features of the energy 
levels of the Dirac electrons as compared to the conventional 
electronic systems are dicussed. We have also evaluated the 
dipole-allowed transitions in the energy levels of the dots. We 
propose that in the high magnetic field limit, the band parameters 
can be accurately determined from the dipole-allowed transitions.
\end{abstract}

\pacs{81.05.Uw,73.63.Kv,73.63.-b}
\maketitle

Quantum dots (QDs), or the `artificial atoms' \cite{qdbook} are 
one of the most intensely studied systems in condensed matter 
physics where the fundamental effects related to various quantum 
phenomena in confined geometries can be studied but with the 
unique advantage that the nature of the confinement and the electron 
density can be tuned externally. However, much of the interest 
on this system derives from its enormous potentials for applications, 
ranging from novel lasers to quantum information processing. 
While the majority of the QD systems investigated are based on the
semiconductor heterostructures, in recent years, quantum dots 
created in the carbon nanotubes have been reported in the literature 
where the `atomic' properties \cite{qd-nanotube1} were clearly 
elucidated and its importance in technological applications 
was also demonstrated \cite{qd-nanotube2}. Conductance properties 
of ultrathin graphitic QDs \cite{graphene-dot} 
have also been reported recently. It is now well recognized that 
the low-energy dynamics of the two-dimensional electrons in 
graphene is governed by the Dirac-Weyl equation, and the charge 
carriers behave as massless chiral fermions \cite{graphene,ando-book}. 
In this situation, confinement of electrons becomes quite a 
challenging task, due to the so-called Klein's paradox 
\cite{geim-klein}. This problem has been dealt with in the 
case of a one-dimensional (1D) wire in zero \cite{efetov} and in
finite \cite{peres} magnetic fields. In this letter, we report on 
the electronic properties of the parabolic QDs in graphene, in particular,
we present the energy levels as a function of the magnetic field
(Fock-Darwin states \cite{qdbook}) and 
the associated dipole-allowed optical transitions in this system. 
We propose that the optical spectroscopy of the graphene QD in the 
high-field limit could provide an accurate means of determining the 
band parameters of graphene.

The Hamiltonian of a single electron in graphene with a
cylindrically symmetric confinement potential is
\begin{equation}
  {\bf H} = {\bf H}_0 + {\bf H}_1 =  \frac\gamma\hbar 
\left( \vec{\sigma} \vec{\pi } \right) +  V(r) ,
\label{H}
\end{equation}
where $\vec{\sigma }$ are the Pauli matrices,
$\vec{\pi} = \vec{p} + \frac ec \vec{A}$, 
$\vec{A} = \frac B2(-y,x)$ is the vector potential corresponding 
to the magnetic field $B$ in the $z$-direction orthogonal to the
graphene plane, and $\gamma = \sqrt3\,a\gamma_0/2$  is the band 
parameter. Here $a=0.246$ nm is the lattice constant and 
$\gamma_0$ (meV) is the transfer integral between the 
nearest-neighbor carbon atoms \cite{saito}. 

At first we analyze the properties of the graphene system in 
the absence of a magnetic field to find the condition for 
confinement of an electron in the potential $V(r)$. Due to the 
Klein tunneling the electrons in graphene can not be localized 
by a confinement potential, since for any potential there will 
be the electron states with negative energy (the hole states) 
which would provide the escape channel for the electron inside 
the potential well. We can then discuss only the quasilocalized 
states or trapping of the electron by the confinement potential. 
This problem has been treated for the quasi-1D graphene system
\cite{falko06,efetov}, where it was shown that the transverse 
momentum in 1D introduces the classically forbidden regions, which 
helps in trapping the electron. The width of the quasilocalized 
level is determined by the tunneling through the classically 
forbidden regions. For the zero transverse momentum, the tunneling 
barriers disappear and there are no trapped states. In our case, 
we have a cylindrically symmetric confinement potential with the 
effective transverse momentum $m/r$, where $m$ is the electron 
angular momentum. Therefore, for $m \neq 0$ we expect the 
trapping of an electron by a cylindrically symmetric QD. In terms
of the two-component wave function $(\chi_1(r) e^{i(m-1)\theta}, 
\chi_2(r) e^{im\theta} )$ the Schr\"odinger equation
corresponding to the Hamiltonian (\ref{H}), is
\begin{eqnarray}
& & V(r) \chi_1 - i \gamma \frac{d\chi_2}{dr} 
-i \gamma \frac mr\chi_2= E\chi_1 
\label{final1}\\
& & V(r) \chi_2 - i \gamma \frac{d\chi_1}{dr} 
+i \gamma \frac{m-1}r\chi_1= E\chi_2. 
\label{final2}
\end{eqnarray}
There are no analytical solution to these equations, and  
%(\ref{final1})-(\ref{final2}). 
therefore, we first present below a semiclassical analysis. 

{\em Semiclassical analysis:} At a large $m$ we 
seek a solution of Eqs. (\ref{final1})-(\ref{final2}) 
in the form $e^{iqr}$, which gives
%$\exp(iqr)$,
% which results in
\begin{equation}
\left( \frac{E-V}{\gamma } \right)^2 = 
\left(\frac{m}{r} \right)^2 + q^2.
\label{q}
\end{equation}
The classical turning points can be found from the condition 
$q=0$, and the classical region is $| E - V(r) | > \gamma |m|/r$. 
If $r_0 $ is the solution of the equation $E-V(r)=0$ then 
we can find the classically forbiden region as 
%\begin{equation}
$(r_0 - \Delta r) < r < (r_0 + \Delta r)$ ,
%\end{equation}
where $\Delta r=m/Fr_0$, $F=\gamma ^{-1} dV(r_0)/dr$, and 
we assumed that $F \gg m/r_0^2$. 
%The electron can classically propogate in the regions $r> r_0 + 
%\Delta r$ and $r < r_0 - \Delta r$. 
If the electron is trapped in the dot, 
i.e., at
%in the region 
$r < r_0 - \Delta r$, 
then the escape rate or the width of the quasilocalized levels is
determined by the tunneling through the classically forbidden region,
%which is given by the following semiclassical expression  
\begin{equation}
T = \exp\left(- \int_{r_0-\Delta r}^{r_0+\Delta r} |q(r)| d r  \right) 
= \exp\left(-\frac{\pi m^2 }{2F r_0^2} \right) .
\label{T}
\end{equation}
Therefore, in order to trap the electron we need a large $m$ and a small $F$, 
i.e. a smooth confinement potential. For a 
potential $V= (u/n) r^n$, Eq.~(\ref{T}) takes the form  
\begin{equation}
T = \exp\left[-\frac{ \pi \gamma m^2}{2 u r_0^{n+1}} \right]= 
\exp\left[- \frac{ \pi m^2 }{n(E/\epsilon _n )^{n+1/n}} \right],
\label{T2}
\end{equation}
where $\epsilon _n = (\gamma ^n u /n)^{1/(n+1)}$. Equation~
(\ref{T2}) gives the upper limit on the energy of the quasilocalized 
levels at a given $m$,
%angular momentum, 
i.e. $E/\epsilon_n < m^{2n/(n+1)}$.   
Based on the semiclassical expression we can also find the 
interlevel separation of the quasilocalized levels 
at large energies, $\Delta E_n =\alpha \epsilon _n (E/\epsilon
_n)^{-1/n}$, where $\alpha \sim 1$. We then 
estimate the number of quasilocalized levels, $N_{n,m}$ for a given 
angular momentum, $m$, and a given potential profile from
\begin{equation}
N_{n,m} = \int^{m^{2n/(n+1)}}_0 \frac{dE}{\Delta E_n}\sim
\frac{n}{n+1} m^2. 
\end{equation}
This estimation is valid for a large $m$. For a small $m$ 
we need to solve the system of equations (\ref{final1})-(\ref{final2}) 
numerically to find the properties of the quasilocalized states. 

\begin{figure}[!ht]
\centerline{\epsfxsize=8.0cm\epsfbox{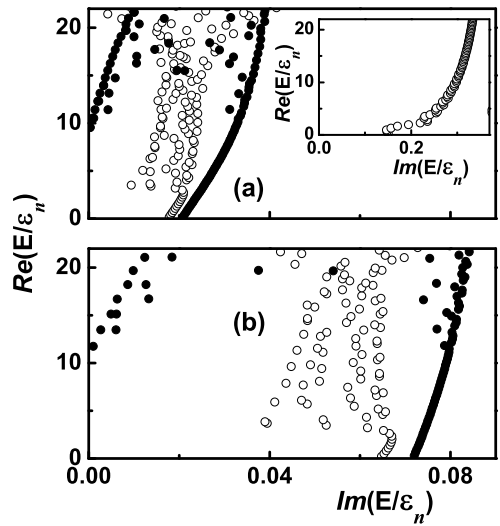}}
\caption[*] {The real and imaginary parts of the energy spectra 
of an electron in a QD with a confinement potential $V(r) = 
(u/n)r^n$, shown for various values of the exponent $n$ and the 
angular momentum $m$: (a) $n=2$, $m=2$  (circles), and   
$n=2$, $m=10$ (dots); (b) $n=4$, $m=2$ (circles), and $n=4$, $m=10$ 
(dots). The results for $n=2$ and $m=1$ are shown as inset. 
The energy is in units of $\epsilon _n$. 
}
\end{figure}

{\em Quasilocalized states -- numerical solutions:} To extract the 
information about the width of the quasibound levels we need to 
impose special boundary conditions far from the QD. This condition 
means that far from the origin, $r\gg r_0$, the solution should be 
an outgoing wave, i.e., the propogation away from the QD. From 
Eqs.~(\ref{final1})-(\ref{final2}), it is clear that for $r\gg r_0$ 
the outgoing solution has the form  
\begin{equation}
\chi_1(r) = -\chi_2(r) = C \exp \left( \frac{i}{\gamma }
\int^{r  } V(r^{\prime} )  dr^{\prime } \right) ,
\label{sol2}
\end{equation}
where $C$ is a constant. Equation (\ref{sol2}) is the boundary 
condition for the system (\ref{final1})-(\ref{final2}) at large 
distances. Since at $r=0$ the solution should be non-divergent,  
another boundary condition is $\chi_1(r=0)=\chi_2(r=0) = 0$. 
A solution with these boundary conditions exists only for a complex 
energy, $E$. The imaginary part of the energy determines the width 
of the quasilocalized level. 

We have solved Eqs.~(\ref{final1})-(\ref{final2}) numerically 
for a potential of the form $V(r)=(\frac un) r^n$ with different 
values of the exponent, $n$, and for different values of the 
angular momentum $m$. In the dimensionless units, i.e., for the 
units of length and the energy, $a_n=(n\gamma /u)^{1/(n+1)}$ and 
$\epsilon _n = (\gamma ^n u /n)^{1/(n+1)}$ respectively, the 
system does not contain any information about the interaction 
strength, $u$. The results in Fig.~1 show a bunch of closely 
spaced levels with a large imaginary part of the energies, 
$Im(E)$. For these states $Im(E)$ is comparable to the interlevel 
spacing. These states are the delocalized contunuum states. At 
the background of the continuum spectra we also see the levels 
with a small imaginary energy, i.e. a small width of the level. 
These states are the quasilocalized states of the QD. The 
manifestation of such states can be seen already at $m=2$ 
(circles) both for $n=2$ [Fig.~1(a)] and $n=4$ [Fig.~1(b)]. 
The strength of the localization can be characterized in 
terms of the ratio of the $Im(E)$ to the interlevel spacing. 
For $m=2$ this ratio is 50. With an increase of the 
magnetic moment the quasilocalized states become well 
developed and at $m=10$ we clearly see the states with 
very low $Im(E)$. The ratio of $Im (E)$ to the interlevel 
spacing for these states is about 800. With an increase of 
the energy the states are less localized, i.e. $Im(E)$ increases. 
This is consistent with Eq.~(\ref{T2}). Note that for all values 
of the exponent $n$ there are no localized states at $m=1$ 
[inset in Fig.~1(a)]. All the states at $m=0$ have very large 
$Im(E)$. There are also no quasilocalized states at $m=0$.
The reason for delocalization of the electron at $m=0$ and $1$ 
is that the effective transverse momentum for either the $\chi_1$ 
component (at $m=0$) or the $\chi_2$ component (at $m=1$) is zero. 
In the following, we analyze the magnetic field effects
on the electronic states of the QDs.

{\em Magnetic Field -- semiclassical analysis:} In a magnetic 
field the system of equations (\ref{final1})-(\ref{final2}) has an
additional non-diagonal term proportional to the magnetic field. 
In the dimensionless units, i.e., for the units of length $a_n$ 
and the energy $\epsilon_n$, the system is characterized by only 
one parameter, $b=(eB/2c)(n\gamma/u)^{2/(n+1)}$. In the 
semiclassical approximation, the effective transverse momentum is 
$(m/r+br)$ and the Eq.~(\ref{q}) becomes
\begin{equation}
T = \exp\left[- \frac{ \pi (m+b\tilde{E}^{2/n})^2 }{
n \tilde{E} ^{n+1/n}} \right],
\label{TH2}
\end{equation}
where $\tilde{E} = E/\epsilon_n$. The effect of the magnetic field 
is different for the states with a positive or a negative $m$ (the 
sign of $m$ depends on the direction of a magnetic field). For 
a positive $m$ the application of a magnetic field increases the 
effective transverse momentum and suppresses the tunneling 
from the QD. For a negative $m$, the magnetic field {\em decreases} 
the tranverse momentum. Therefore the state becomes less 
localized. If we increase the magnetic field even further then 
at some point, $b = m/\tilde{E}^{2/n}$, the level becomes 
{\em delocalized}, and at a even larger $B$ the level again 
becomes localized. Now the trapping will be due to the magnetic 
field. Therefore for a negative $m$, the magnetic field induces a
{\em localization-delocalization-localization} transition.

The number of the quasilocalized states in a weak magnetic field 
is estimated to be $N_{n,m} \sim \left[m+b|m|^{2/(n+1)} 
\right]^2$. This number with a positive $m$ increases with an 
increasing magnetic field, while that for a negative $m$ decreases 
with the magnetic field upto a certain value of $B$ and then 
increases. The total number of states with positive and negative 
angular momenta, $N_{n,m}+N_{n,-m} \sim m^2 + b^2|m|^{4/(n+1)}$
always increases with an increasing $B$. From this bahavior we 
expect the following effect: We assume that the QD is occupied 
by electrons upto a certain energy, i.e., the states with both 
positive and negative angular momenta are occupied and the net 
angular momentum of the dot is zero. We then apply a magnetic 
field and the states with positive $m$ becomes more localized while 
the electrons from the states with negative $m$ will escape from the 
QD. Finally, the electrons in the QD will have a net 
positive angular momentum and correspondingly a net magnetic moment.
 
{\em Magnetic Field -- numerical results:} To study the dependence 
of the quasilocalized spectra on the magnetic field we introduce 
the wavefunctions of the Hamiltonian ${\bf H}_0$, i.e. without a 
confinement potential \cite{zheng}, as the basis functions. To 
eliminate any escape of the electron from the QD we 
consider only the basis functions with the positive energy,
\begin{equation}
  \Psi_{n,m} = C_N \left(
  \begin{array}{c}
    sgn(N) \phi_{n,m-1}(x) \\
    i \phi_{n,m}(x)
  \end{array} \right) \;,
\end{equation}
where $N=n+\frac12(|m|+m)$ is the Landau Level (LL) index, 
$C_{N=0}=1$ and $C_{N\neq 0} =1/\sqrt2$,  
$sgn(N=0)=0$, 
and
\begin{equation}
  \phi_{n,m} = \frac1{\sqrt{2}} \sqrt{\frac{n!}{(n+|m|)!}} 
    e^{-x/2} x^{|m|/2} L_n^{|m|}(x) e^{im\theta}
\end{equation}
is the Landau wavefunction. Here $L_n^{|m|}(x)$ is the 
associated Laguerre polynomial, $x=r^2/{a^\prime}^2$ is a 
dimensionless distance. Here ${a^\prime}$ is the characteristic 
length of the system. Without the confinement ${a^\prime}$ 
should be equal to the mangetic length $l=\sqrt{\hbar c/eB}$. 
In the presence of the confinement, the Hamiltonian suggests 
a natural unit of length $(\gamma/u)^{1/3}$ \cite{efetov} and 
a natural unit of energy $(\gamma^2u)^{1/3}$. This length 
characterizes the size of a parabolic dot in graphene. 
Therefore, 
$$ \frac2{{a^\prime}^2}=\frac1{l^2}+\left( \frac{u}\gamma
\right)^{2/3}.$$

\begin{figure}[!ht]
\centerline{\epsfxsize=8.0cm\epsfbox{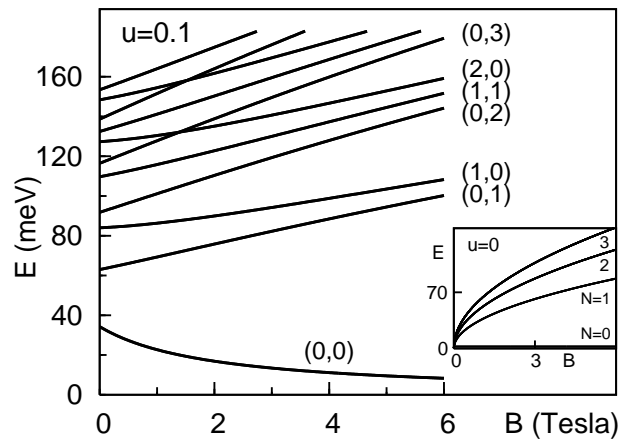}}
\caption[*] {Fock-Darwin spectrum of the Dirac quantum dots, 
plotted for the confinement potential strength $u=0.1$ (in
dimensionless units).
The numbers in the parentheses correspond to the two quantum 
numbers $n$ and $m$. Results for $u=0$ are given as an inset.
}
\end{figure}

In our numerical calculations, we choose the band parameter 
to be $\gamma=646$ meV$\cdot $nm for $\gamma_0=3.03$ eV 
\cite{ando_review}. The low-lying energy states of the 
graphene QD are shown in Fig.~2. In the absence of a 
confinement potential, the Dirac spectrum scales as 
$\sqrt2\,\gamma\sqrt{N}/l$ [shown as inset]. In the Fock-Darwin 
spectrum for a conventional electron dot, the energy levels 
are degenerate and equally spaced at $B=0$ \cite{qdbook}. 
The two-dimensional parabolic confinement considered here 
shows two outstanding features in contrast to the Fock-Darwin 
spectra at $B=0$. The first is the lifting of the degeneracy 
and the other is the unequal seperation among the energy levels. 
Figure 2 shows the field-dependent energy spectrum for $u=0.1$. 
The energy difference between the lowest two levels at $B=0$ is 
about $(\gamma^2u)^{1/3}$. At a low magnetic field, the 
magnetic length $l$ is larger than or comparable to the size of 
the confinement $(\gamma/u)^{1/3}$ and there is a hybridization 
of the LLs with the levels arising from the spatial confinement. 
In the high magnetic field limit $l \ll (\gamma/u)^{1/3}$, 
the Landau-type levels prevail, as expected. The Fock-Darwin 
spectra for conventional quantum dots have been determined 
earlier by the transport spectroscopy \cite{fock_expt}. Similar 
studies for the graphene QDs would be very important to explore 
the energy levels and the nature of confinement for Dirac
fermions in a graphene QD.

\begin{figure}[!th]
\centerline{\epsfxsize=8.0cm\epsfbox{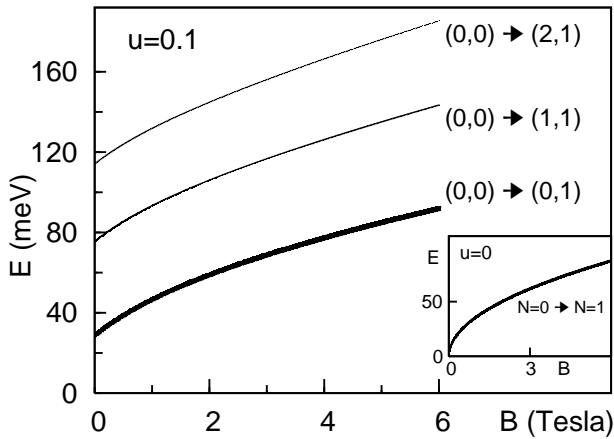}}
\caption[*]{The dipole-allowed transitions in the Fock-Darwin 
spectrum of graphene QDs for $u=0.1$. Inset: the case of $u=0$.
The thickness of the lines is proportional to the calculated 
intensity. From the bottom to the top, the relative intensities 
are about 1.0, 0.1, 0.02, respectively.}
\end{figure}

Figure 3 shows the dipole-allowed optical absorption spectra 
\cite{qdbook,heitmann} for $u=0$ and $u=0.1$. Without any 
confinement there is only the $(0,0)\rightarrow(0,1)$ transition 
(shown as inset in Fig.~3). The additional transitions, 
$(0,0)\rightarrow(1,1)$ and $(0,0)\rightarrow(2,1)$ are due 
to the presence of the parabolic confinement. For the Dirac 
fermions in a QD, the lowest dipole-allowed transition is
\begin{equation}
  \Delta E \simeq \left( \gamma^2 u \right)^{1/3} 
  + \frac{\sqrt2\,\gamma}l.
\end{equation}
For $B=0$ the corresponding energy is $\approx(\gamma^2u)^{1/3}$.  
As the magnetic field increases, $\Delta E$ approaches the 
cyclotron energy $\sqrt2\,\gamma/l = \sqrt2\, (e\gamma B/\hbar c)$,
where $\gamma$ can therefore be uniquely determined experimentally.  
Conventionally, in the nearest-neighbor tight-binding model, 
$\gamma_0$ is obtained by fitting the {\it ab initio} calculation 
and the experimental data \cite{saito}.  In a GaAs quantum dot, the 
magnetic-field-dependent far-infrared absorption 
experiments have established the energy relation $\Delta E_\pm 
= \hbar\Omega \pm \frac12 \hbar\omega_c$ to a great accuracy
\cite{heitmann}.  Similarly, for the massless chiral fermions
in a graphene QD, we expect that the band parameter 
$\gamma$ can also be determined quite accurately by the optical 
absorption experiments in the high-field limit.

Although the high magnetic field results are the major focus 
of this paper as in this case the localization of the electron 
in a QD at all values of the $m$ is provided by the magnetic 
field, at B=0 the chiral nature of the states prevent the electrons 
from being confined in a QD. This clearly indicates that the nature 
of the energy states and the optical spectra at a very small $B$ 
are still important open questions.

We would like to thank P. Pietil\"ainen and Xue-Feng Wang 
for helpful discussions. The work has been supported by the 
Canada Research Chair Program and a Canadian Foundation for 
Innovation Grant.


\begin{thebibliography}{99}

\bibitem{qdbook}
T. Chakraborty, {\it Quantum Dots} (Elsevier, Amsterdam, 1999);
T. Chakraborty, Comments Condens. Matter Phys. {\bf 16}, 35
(1992). 

\bibitem{qd-nanotube1}
M.R. Buitelaar et al., 
%A. Bachtold, T. Nussbaumer, M. Iqbal, and C. Sch\"onenberger, 
Phys. Rev. Lett. {\bf 88}, 156801 (2002); 
D.H. Cobden, and J. Nyg\aa rd, {\it ibid.} {\bf 89}, 046803 
(2002); S. Moriyama et al., 
%T Fuse, M. Suzuki, Y. Aoyagi, and K. Ishibashi, 
{\it ibid.} {\bf 94}, 186806 (2005); S.-H. Ke, 
H.U. Baranger, and W. Yang, {\it ibid.} {\bf 91}, 116803 
(2003).

\bibitem{qd-nanotube2}
K. Ishibashi et al., 
%S. Moriyama, D. Tsuya, and T. Fuse, 
J. Vac. Sci. Technol. A {\bf 24}, 1349 (2006).

\bibitem{graphene-dot}
J. Scott Bunch et al., 
%Y. Yaish, M. Brink, K. Bolotin, and P.L. McEuen, 
Nano Lett. {\bf 5}, 287 (2005).

\bibitem{graphene}
K.S. Novoselov et al., Nature {\bf 438}, 197 (2005); 
Y. Zhang et al, {\it ibid.} {\bf 438}, 201 (2005).

\bibitem{ando-book}
T. Ando, in {\it Nano-Physics {\&} Bio-Electronics: A New 
Odyssey}, edited by T. Chakraborty, F. Peeters, and U. Sivan 
(Elsevier, Amsterdam, 2002), Chap. 1.

\bibitem{geim-klein}
M.I. Katsnelson, K.S. Novoselov, and A.K. Geim, Nature Phys. 
{\bf 2}, 620 (2006);
O. Klein, Z. Phys. {\bf 53}, 157 (1929); {\bf 41}, 407 (1927);
A. Calogeracos and N. Dombey, Contemp. Phys. {\bf 40}, 313 (1999).

\bibitem{efetov}
P.G. Silvestrov and K.B. Efetov, Cond-mat/0606620 (unpublished).

\bibitem{peres}
N.M.R. Peres, A.H. Castro Neto, and F. Guinea, Phys. Rev. B 
{\bf 73}, 241403 (2006).

\bibitem{saito}
R. Saito, G. Dresselhaus, and M.S. Dresselhaus, {\it Physical
Properties of Carbon nanotubes} (Imperial College Press, 
London, 1998).

\bibitem{falko06} V. V. Cheianov and V. I. Falko, 
Phys. Rev. B {\bf 74}, 041403 (2006)

\bibitem{zheng}
Y. Zheng and T. Ando, Phys. Rev. B {\bf 65}, 245420 (2002).

\bibitem{ando_review}
T. Ando, J. Phys. Soc. Jpn. {\bf 75}, 074716 (2006).

\bibitem{fock_expt}
P.L. McEuen et al., Phys. Rev. Lett. {\bf 66}, 1926 (1991); 
J. Weis et al., 
%R.J. Haug, K. v. Klitzing, and K. Ploog, 
Phys. Rev. B {\bf 46}, 12837 (1992); T. Schmidt et al., {\it ibid.} 
{\bf 51}, 5570 (1995); S. Tarucha et al., Phys. Rev. Lett. 
{\bf 77}, 3613 (1996).

\bibitem{heitmann}
C. Sikorski and U. Merkt, Phys. Rev. Lett. {\bf 62}, 2164 
(1989); B. Meurer, D. Heitmann, and K. Ploog, Phys. Rev. 
Lett. {\bf 68}, 1371 (1992); D. Heitmann and J. Kotthaus, 
Phys. Today {\bf 46}, 56 (1993).

\end{thebibliography}
\end{document}